# New approach for the molecular beam epitaxy growth of scalable single-crystalline WSe$_2$ monolayers

C. Vergnaud[*], M.-T. Dau, A. Marty, M. Jamet
Univ. Grenoble Alpes, CEA, CNRS, Grenoble INP, IRIG-Spintec, 38000 Grenoble, France

B. Grévin
Univ. Grenoble Alpes, CNRS, CEA, IRIG-SYMMES, 38000 Grenoble, France

H. Okuno
Univ. Grenoble Alpes, CNRS, CEA, IRIG-MEM, 38000 Grenoble, France

C. Licitra
Univ. Grenoble Alpes, CEA, LETI, Minatec Campus, 38000 Grenoble, France

**Abstract:**

The search for high-quality transition metal dichalcogenides mono- and multi-layers grown on large areas is still a very active field of investigation nowadays. Here, we use molecular beam epitaxy to grow 15×15 mm large WSe$_2$ on mica in the van der Waals regime. By screening one-step growth conditions, we find that very high temperature (>900°C) and very low deposition rate (<0.015 nm/min) are necessary to obtain high quality WSe$_2$ films. The domain size can be larger than 1 µm and the in-plane rotational misorientation less than ±0.5°. The WSe$_2$ monolayer is also robust against air exposure, can be easily transferred over 1 cm$^2$ on SiN/SiO$_2$ and exhibits strong photoluminescence signal. Moreover, by combining grazing incidence x-ray diffraction and transmission electron microscopy, we could detect the presence of few misoriented grains. A two-dimensional model based on atomic coincidences between the WSe$_2$ and mica crystals allows us to explain the formation of these misoriented grains and gives suggestion to remove them and further improve the crystalline quality of WSe$_2$.

---

[*] Email address : celine.vergnaud@cea.fr

**Introduction**

In the past decade, layered transition metal dichalcogenides (TMDs) have attracted tremendous attention in the scientific community for their exceptional electrical and optical properties. Depending on the transition metal and chalcogen atoms, their electronic properties can range from semiconducting, to metallic and superconducting[1,2]. The most popular semiconducting TMDs are $MX_2$ with M=Mo or W and X=S or Se because they exhibit an indirect-to-direct band gap crossover when thinned down to a single layer[3]. This unique transition leads to very high photoluminescence emission with strong excitonic effects[4]. Owing to the crystal inversion symmetry breaking and strong spin-orbit interaction, another key property of $MX_2$ single layers is the presence of two distinct valleys K+ and K- with opposite spin-splitting in the two-dimensional Brillouin zone[5]. Addressing and manipulating this new valley degree of freedom gives rise to a new field of investigation called valleytronics[6,7] that might be very promising for future electronic devices. In the $MX_2$ family, $WSe_2$ is of particular interest for several reasons: a very large spin-orbit coupling (>400 meV) and spin splitting in the K-valleys[8], a long valley coherence[9], an ambipolar behaviour when processed into a transistor[10], and a rather high carrier mobility (~250 $cm^2$/V s)[11,12].

To date, most of the physical studies have been performed on micron-sized $WSe_2$ flakes mechanically exfoliated from the bulk. Efforts to grow large-area high-crystalline quality $WSe_2$ layers using bottom-up approaches such as chemical vapour deposition (CVD)[13,14], metal-organic CVD[15,16] or molecular beam epitaxy MBE[17,18,19,20,21,22] are still ongoing. In order to anticipate the fabrication of electronic devices, the substrate on which to grow $WSe_2$ has to be of low cost, abundant, available in large areas, and promote the growth of single crystalline films. The single-crystalline character insures large carrier mobilities and optical response by reducing the density of grain boundaries and scattering processes. Finally, the substrate should be insulating to well define the conduction channel. By processing the TMD layers directly on the insulating substrate, one avoids any layer transfer process that degrades the electronic properties of TMD layers by the use of water, organic solvents or polymers. The MBE growth of WSe$_2$ on highly oriented pyrolytic graphite (HOPG) or epitaxial graphene prevents the study of electrical and optical properties because the substrate is conducting and is quenching the photoluminescence (PL) by fast charge transfer[23]. In Ref. 11, WSe$_2$ was grown layer-by-layer on sapphire which is more adapted to the study of physical properties for its insulating character. Despite the demonstration of ambipolar carrier transport when processed into a transistor, the PL of WSe$_2$ monolayer was not shown and the electrical measurements were performed on multilayers and not on a single layer as required for valleytronics applications. Moreover, in all previous works, the MBE growth was performed in two steps: the co-evaporation of W and Se (Se being in large excess with respect to W) at a moderate temperature (<500°C) to ensure the efficient incorporation of Se in the film and the formation of the stoichiometric compound followed by an annealing at higher temperature (>600°C) under Se to improve the crystalline quality. This two-step method provides highly uniform TMD single or multi layers over large areas with almost 100 % surface coverage. In addition, when the substrate is single-crystalline (like epitaxial graphene or sapphire), the WSe$_2$ film shows a clear in-plane texture as a consequence of the van der Waals epitaxy[17,21]. However the grains remain small (<200 nm) and exhibit an in-plane rotational misorientation larger than ±5° with respect to the substrate crystal axes. The small grain size along with the grain-to-grain rotational misorientation results in the formation of defective grain boundaries that may affect the air stability[24], carrier mobility and optical properties of WSe$_2$. However, if all the WSe$_2$ monolayer crystallites are well-oriented and coalesce at the scale of the substrate (cm) during the growth, they can form large scale high quality single crystal domains[25].

In this communication, we develop a one-step MBE growth method to grow WSe$_2$ by van der Waals epitaxy on mica. Mica complies with all the requirements for device fabrication:

low cost, abundant, insulating and available in large areas. A broad range of growth temperatures and W deposition rates are explored to minimize in-plane rotational misorientation and to increase the final domain size. Muscovite mica [$KAl_2(AlSi_3O_{10})(OH)_2$] is a lamellar monoclinic material made of silicate sheets held together by a layer of potassium atoms[26]. It is electrically insulating, single-crystalline and exhibits atomically flat terraces over hundreds of microns. Moreover, the absence of dangling bonds at the surface of mica makes it an ideal substrate to test the van der Waals epitaxy of $WSe_2$. We also demonstrated the transfer of 1 $cm^2$ TMD layers from mica onto a $SiO_2$/Si substrate by dipping the stack into deionized water and using the hydrophilic character of the mica surface[27]. The mica substrate is then reusable to grow another $WSe_2$ layer. Here we show that for the highest growth temperature (920°C) and lowest W flux (F=0.15 Å/min), the in-plane rotational misorientation or twist is ±0.5° i.e. one order of magnitude smaller than in previous reports. The high growth temperature also favors the re-crystallization at the grain boundaries. As a consequence we can detect an intense PL signal of $WSe_2$ on mica at room temperature after several days of air exposure. This is the first time such a strong and stable PL signal is measured on MBE-grown $WSe_2$ monolayers. Moreover, by varying the growth conditions, we could elucidate the origin of the in-plane rotational misorientation as a direct consequence of the van der Waals epitaxy mechanism. This work provides a new and comprehensive route to grow $WSe_2$ monolayers compatible with optoelectronics or valleytronics applications.

**Experimental**

The 15×15 $mm^2$ mica substrate from Ted Pella Inc. is first exfoliated mechanically with scotch tape to produce a fresh and atomically smooth surface prior to its introduction into the ultrahigh vacuum (UHV) system with a base pressure of $3\times10^{-10}$ mbar. It is then outgassed in the MBE chamber at 700°C for 5 minutes. $WSe_2$ layers are grown by co-evaporating W from an e-gun evaporator and Se from a standard Knudsen cell. A quartz micro-balance regulates the W flux and the Se pressure is fixed at $1\times10^{-6}$ mbar as given by a retractable Bayard-Alpert gauge in place of the sample. In-situ reflection high energy electron diffraction (RHEED) is used to monitor the $WSe_2$ crystal structure during the growth. Muscovite mica being a sheet silicate, it consists of layered silicon-oxygen tetrahedra. The sheets are 1 nm-thick and weakly linked together by a layer of potassium atoms. The weakest bonds in the crystal are those between potassium and oxygen so that cleavage takes place along a potassium layer. After exfoliation, the mica surface is thus made of a silicon-aluminum oxide layer half covered by randomly distributed potassium atoms[26].

In order to produce the best crystalline quality $WSe_2$ layers, we vary the W flux from F=0.15 Å/min to F=5.25 Å/min and the growth temperature from T=520°C to T=920°C. To monitor the crystal quality, we use two kinds of measurements: Raman spectroscopy and grazing incidence x-ray diffraction. Raman spectra of $WSe_2$ layers are recorded using a confocal microscope with a 632.8 nm excitation wavelength and a Horiba LabRAM HR spectrometer. In non-resonant conditions, the Raman spectrum is usually dominated by the vibration modes lying at the Γ point. Here, because the excitation wavelength is close to the energy gap of $WSe_2$, the spectrum is observed in resonant Raman scattering conditions and the observed peaks are identical to those observed by Bhatt et al.[28] as shown in the inset of Fig. 1a. X-ray diffraction measurements are performed with a SmartLab Rigaku diffractometer. The source is a copper rotating anode beam tube ($K_\alpha$ = 1.54 Å) operating at 45 kV and 200 mA. The diffractometer is equipped with a parabolic multilayer mirror and in-plane collimators of 0.15° (resp. 0.114°) on the source (resp. detector) sides defining the angular resolution. A $K_\beta$ filter on the detector side eliminates parasitic radiations. We consider three figures of merit to assess the crystalline quality of $WSe_2$ monolayers: the full width at half maximum (FWHM) of the $A_{1g}$ out-of-plane vibration mode of $WSe_2$ in Raman spectra $\Delta A_{1g}$, the FWHM of the (100) diffraction line along a radial scan $\Delta\theta$ (reflecting the grain size parallel to (100) and distribution of lattice parameters) and an azimuthal scan $\Delta\phi$ (reflecting the grain size perpendicular to (100) and in-plane rotational misorientation). Examples are shown in the insets of Fig. 1a, 1b and 1c respectively. In the Raman spectrum, since several peaks overlap ($E^1_{2g}$-$E^2_{2g}$, 2LA-$E^2_{2g}$, $E^1_{2g}$, $A_{1g}$ and 2LA) due to the resonant excitation conditions, a fitting procedure is necessary to extract the FWHM of the $A_{1g}$ peak. Concerning x-ray diffraction, we systematically plotted and fitted the $\delta q^2(q^2)$ curves where q is the scattering vector for (100), (110) and (200) Bragg peaks in the radial direction to deduce an average grain size[29]. We find a lower bound (because of the instrumental resolution) of 30 nm. This grain size cannot justify the large FWHM of the peaks in azimuthal scans which mainly reflects the in-plane rotational misorientation of WSe$_2$ grains. Figure 1 summarizes the results for the growth of 1 ML of WSe$_2$. The diameter of the blue, red and yellow spheres is proportional to the value of $\Delta A_{1g}$, $\Delta\theta$ and $\Delta\phi$ respectively. The best crystalline quality corresponds to the lowest values of these parameters. They all three evolve identically with the growth parameters: they are decreasing when increasing the growth temperature and decreasing the W flux. We conclude that the best crystalline quality is obtained for T=920°C and a W flux F=0.15 Å/min. This growth temperature is much higher than the one conventionally used for the MBE-growth of TMDs and is very close to the sublimation point of WSe$_2$ of ~940°C[30]. It is also close to the delaminating temperature of mica close to 950°C. F=0.15 Å/min corresponds to the minimum stable flux we can reach with our e-gun evaporator. For (T=920°C, F=0.3 Å/min), we find the following parameters: $\Delta A_{1g}$= 5.3 cm$^{-1}$, $\Delta\theta$=0.33° and $\Delta\phi$=1.25°.

As a comparison, we show in Fig. 2 the RHEED patterns, atomic force microscopy (AFM) images and the azimuthal scan of the (100) diffraction peak of WSe$_2$ for two different samples grown at (T=720°C, F=1.5 Å/min) and (T=920°C, F=0.3 Å/min). AFM images are acquired in the peak force mode of a Bruker's Dimension Icon microscope. By increasing the growth temperature and decreasing the W flux, the RHEED patterns show a clear anisotropic character whereas they exhibit an isotropic character for low temperature and high flux. Moreover, the grain size increases from 10 nm (this small size giving rise to the granular aspect of the AFM image in Fig. 2c) up to 100 nm and the grains morphology evolves from round to triangular shape. Finally, the in-plane rotational misorientation decreases from 14.60° down to 1.25° which is comparable to standard three-dimensional semiconductor epitaxy like aluminium-gallium nitrides[31]. In the following, we discuss in detail about the structural and optical properties of samples grown at high temperature and low flux.

**Results and discussion**

In Fig. 3, we show the full set of grazing incidence x-ray diffraction scans and we discuss about the peak positions and width. First, the Bragg peak positions in Fig. 3a correspond to a lattice parameter of 3.28 Å which is the one of bulk $WSe_2$. In Fig. 3b, the most intense peaks are detected every 60° (numbered 1 in the yellow scan) demonstrating that $WSe_2$ grains are in epitaxial relationship with the mica substrate despite the large lattice mismatch of 10 %. The epitaxial growth of $WSe_2$ on mica is made possible by the weak van der Waals interaction between the epilayer and the substrate. We also observe several satellite peaks that could be reproducibly observed in several samples grown in the same conditions. The corresponding azimuthal angles are: $\delta\phi=11°$ (peak number 6), $\delta\phi=18°$ (peak number 4), $\delta\phi=30°$ (peak number 3), $\delta\phi=42°$ (peak number 2) and $\delta\phi=49°$ (peak number 5). Considering the $WSe_2$ crystal symmetries, peaks 2 and 4 (resp. 5 and 6) are equivalent. As shown in Fig. 3c, the relative integrated intensity of peak 1, peak 3, peak 2-4, and peak 5-6 are 74 %, 10 %, 4×2=8 % and 4×2=8 % respectively. We attribute those peaks to the presence of highly misoriented grains.

In order to confirm their presence, we carried out scanning transmission electron microscopy-high-angle annular dark field (STEM-HAADF) observations and 4D STEM diffraction. The 4D STEM diffraction maps are acquired using the Gatan STEMX system. Images are processed using the software packages of Gatan Microscopy Suite 3.0. For these observations, the same $WSe_2$ film is transferred from mica onto a TEM copper grid using a wet transfer process. The $WSe_2$/mica stack covered by a thin polystyrene film is dipped into de-ionized water. Due to the hydrophilic character of the mica surface, water infiltrates between the $WSe_2$ layer and mica which forces $WSe_2$ to detach from the substrate and float at the surface of water. Using a fishing procedure, the $WSe_2$ layer is then transferred onto a TEM copper grid and polystyrene removed by acetone[27]. In Fig. 4a, we show the STEM-HAADF image in-plane view of the transferred layer which corresponds exactly to the AFM image in Fig. 2c with the presence of triangular grains. This observation first confirms the successful wet transfer of the $WSe_2$ layer to the TEM grid. By using 4D STEM, we are able to resolve individual grain orientation in Fig. 4b. In agreement with x-ray diffraction, we find that a majority of grains (coloured in yellow in Fig. 4b) exhibit the same crystal orientation. They correspond to the grains in epitaxial relationship with mica and a corresponding electron diffraction pattern is shown in Fig. 4c and numbered 1 by analogy with the Bragg peaks in Fig. 3b. Those grains eventually coalesce to form micron-sized grains with in-plane rotational misorientation ±0.5°. In Fig. 4b and 4c, we observe the presence of $WSe_2$ grains rotated by +38°, +30° and +17.5° with respect to the mica orientation. In Fig. 4b, those grains are coloured in blue, red and green respectively and the electron diffraction patterns are displayed in Fig. 4c. The orientations correspond well to the satellite Bragg peaks of Fig. 3b and are numbered 2; 3 and 4 accordingly. The individual dark field images for the 4 crystal directions are shown in Fig. 4d. The grains corresponding to the peaks 5 and 6 could also be observed in other STEM images. To summarize, we have confirmed at the microscopic scale the existence of misoriented WSe$_2$ grains observed by x-ray diffraction at the macroscopic scale. In order to get more insight into the van der Waals epitaxy of WSe$_2$ on mica and justify the existence of misoriented grains, we develop a simple model based on the two-dimensional atomic coincidences between the WSe$_2$ and mica lattices. As shown in Fig. 5a (cross section) and 5b (top view), we start with a sheet of mica and put a triangle of WSe$_2$ on top at an arbitrary gap value. The size of the triangle is 3 nm (≈10 lattice parameters) and corresponds to the one observed by Yue et al.[18] at the very first stage of WSe$_2$ growth on HOPG. We can thus consider that it is the first and smallest WSe$_2$ crystallite to form at the surface of mica.

Contrary to conventional epitaxy, in the van der Waals epitaxy, the atomic positions of the WSe$_2$ crystallites are not locked to those of the mica substrate. This crystallite is thus free to translate or rotate at the surface of mica. During the film deposition, the crystallite will grow and reach a critical size above which the atomic positions are frozen. At the critical size, the energy cost to translate or rotate the crystallite is too high with respect to k$_B$T. When the WSe$_2$ crystallite is free to rotate at the surface of mica, it will spend more time at positions that maximize the overlap between the WSe$_2$ and mica atomic orbitals at the interface to minimize the system energy. In other words, these positions maximize the atomic coincidences between the two lattices which gives rise to the so-called Moiré patterns. In our model, the parameter to "measure" the atomic coincidences is the sum of all the distances between each atom of the WSe$_2$ crystallite and its nearest neighbour in mica. For the sake of simplicity, we only consider W atoms from the crystallite and Al, Si and K atoms from the topmost mica sheet. The sum is carried out in two-dimensions: only in-plane distances are considered, we ignore out-of-plane atomic distances and the van der Waals gap. The WSe$_2$ crystallite is then rotated from δϕ=0° to δϕ=120° on the fixed mica lattice around one W central atom which is arbitrarily located on top of one Al, Si or K atom. The sum is then calculated for each δϕ angle. Atomic coincidences correspond to the minima of this sum. The results are shown in Fig. 5c for W-K distances (purple), W-Al distances (dark blue) and W-Si distances (light blue) respectively. Each curve exhibits minima at: $\delta\phi=0°$, 30° and 60° for W-Al and W-Si and $\delta\phi=0°$, 11°, 22°, 38°, 49° and 60° for W-K.

Interestingly, the minima positions match the peak positions of x-ray diffraction data and STEM observations (except peaks 2 and 4 shifted by ≈4°). The corresponding numbers are indicated in Fig. 5c. This result provides a qualitative explanation for the experimental orientation of $WSe_2$ grains. Peaks 1 correspond to $WSe_2$ grains in epitaxy on either the Al, Si or K lattices, peak 3 to $WSe_2$ grains in epitaxy on the Al or Si lattice and peaks 2; 4; 5 and 6 to $WSe_2$ grains in epitaxy on the K lattice. From x-ray diffraction data in Fig. 3c, we can conclude that at least 16 % of the $WSe_2$ grains have grown in epitaxy on K-rich areas: 4 % for each grain orientation (11°, 18°, 42° and 49°). Therefore, in order to improve the $WSe_2$ crystalline quality on mica, K atoms should be removed either physically or chemically from the mica surface. By this, only grains at $\delta\phi=0°$ and 30° will be present at the surface reducing the density of grain boundaries and improving the electrical and optical properties of the film.

Finally, to benchmark the quality of these $WSe_2$ films grown on mica, we measure the photoluminescence of one monolayer of $WSe_2$ grown at (T=920°C, F=0.15 Å/min). At the same growth temperature, by decreasing the W flux from 0.3 Å/min down to 0.15 Å/min, we favour the two-dimensional growth and obtain high surface coverage and very big $WSe_2$ grains as shown in Fig. 6a by AFM. In this sample, the monolayer grain size can reach 1 µm. We then perform photoluminescence measurements at room temperature on this sample after one week of air exposure. The PL spectrum is shown in Fig. 6b along with the one of the same film transferred on SiN(10 nm)/$SiO_2$(300 nm)/Si. First, the PL intensity on mica is comparable to the one obtained on monolayer flakes[32] but on 1 $cm^2$. Moreover, the PL intensity is enhanced by a factor 6 for the transferred film due to optical interferences into the $SiO_2$ layer. These results reflect the high crystalline quality of $WSe_2$ films as well as their robustness and stability in air.

**Conclusion**

In conclusion, we have shown in this work that high growth temperatures and very low W flux are necessary to grow high quality single-crystalline $WSe_2$ films down to one monolayer by van der Waals epitaxy on mica. Using these growth conditions, we find that a majority (75 %) of the $WSe_2$ grains are in exact epitaxial relationship with mica exhibiting an in-plane rotational misorientation as low as ±0.5°. We also detect the presence of few misoriented grains (by 11°, 18°, 30°, 42° and 49° with respect to the mica crystal) both macroscopically by x-ray diffraction and microscopically by transmission electron microscopy. Using a simple two-dimensional model of atomic coincidence between the $WSe_2$ and mica atoms, we demonstrate that the presence of those grains is due to the van der Waals epitaxy growth mechanism itself. Based on this model, we also propose solutions to improve the crystalline quality of $WSe_2$ epiaxially grown on mica. For high growth temperature and very low deposition rate, we eventually detect a strong photoluminescence signal after long air exposure that is even enhanced after the layer is transferred on $SiN/SiO_2$. It shows the high quality of the epitaxial films, their robustness and stability in air.

**Conflicts of interest**

There are no conflicts to declare.

**Acknowledgements**

This research was funded by the Agence nationale de la recherche ANR-MAGICVALLEY ANR-18-CE24-0007. The LANEF framework (ANR-10-LABX-5101) is also acknowledged for its support with shared infrastructure. The authors would like to acknowledge the IRIG/SYMMES/STEP laboratory of the CEA-Grenoble for Raman spectroscopy.

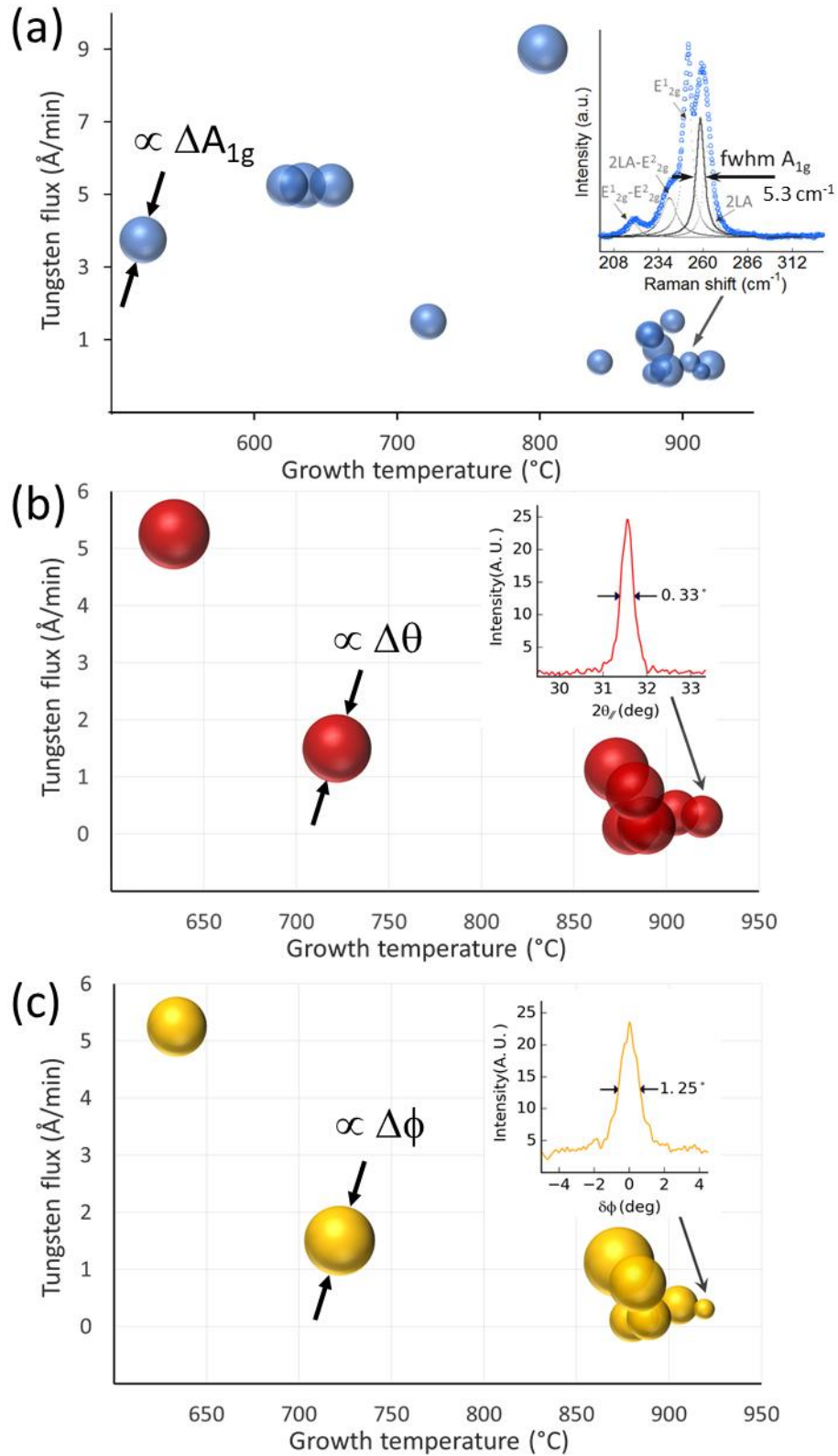

Figure 1: Evolution of the FWHM of the $A_{1g}$ Raman peak $\Delta A_{1g}$ (a), (100) radial diffraction peak $\Delta\theta$ (b) and azimuthal diffraction peak $\Delta\phi$ (c) as a function of the growth temperature and W flux. The insets show examples of Raman and diffraction spectra for specific growth conditions.

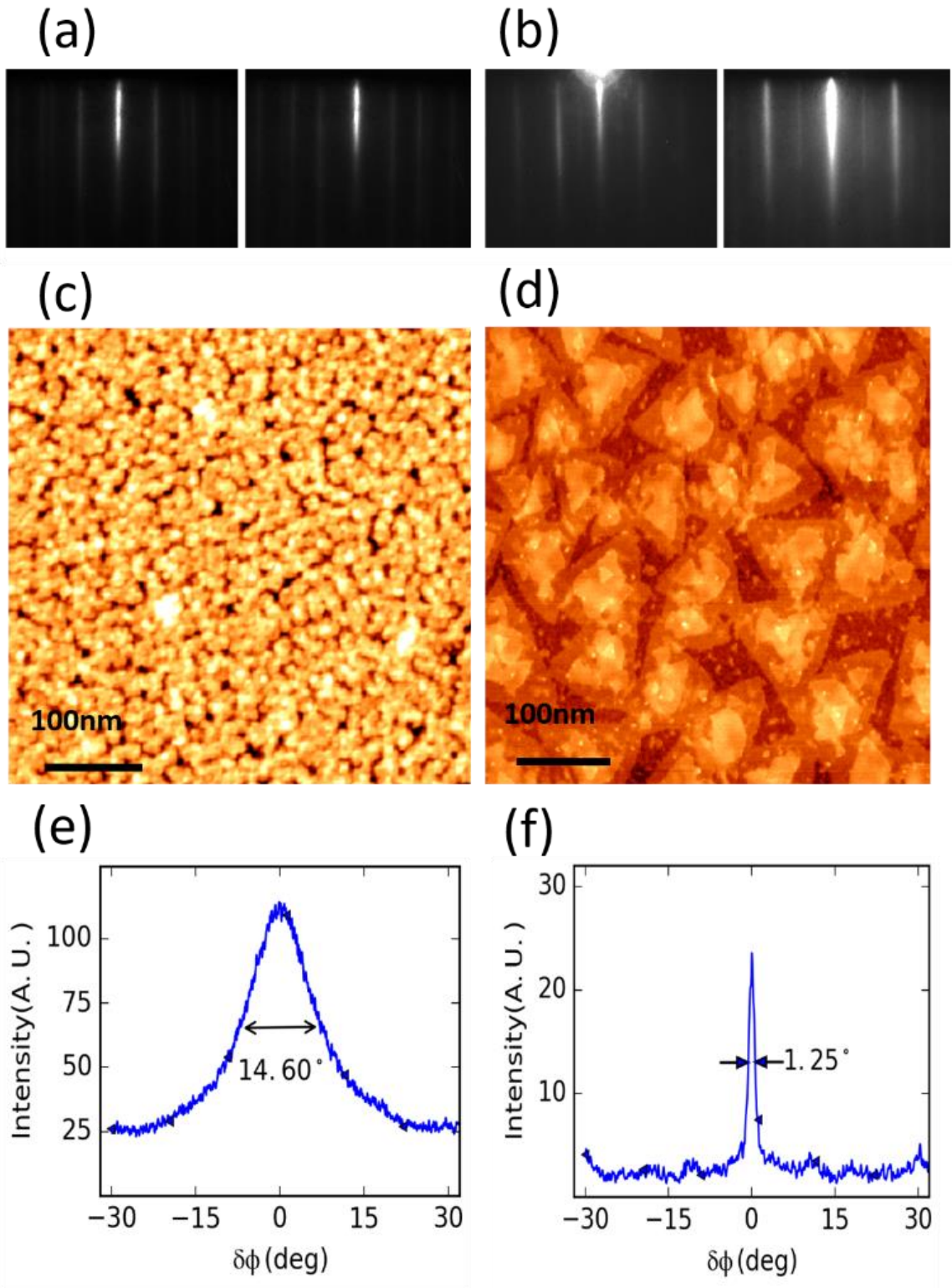

Figure 2: (a), (b) RHEED patterns along (100) and (110) directions. (c), (d) AFM images and (e), (f) azimuthal grazing incidence x-ray diffraction spectra showing the (100) peak FWHM for a WSe$_2$ monolayer grown at (T=720°C, F=1.5 Å/min) and (T=920°C, F=0.3 Å/min) respectively.

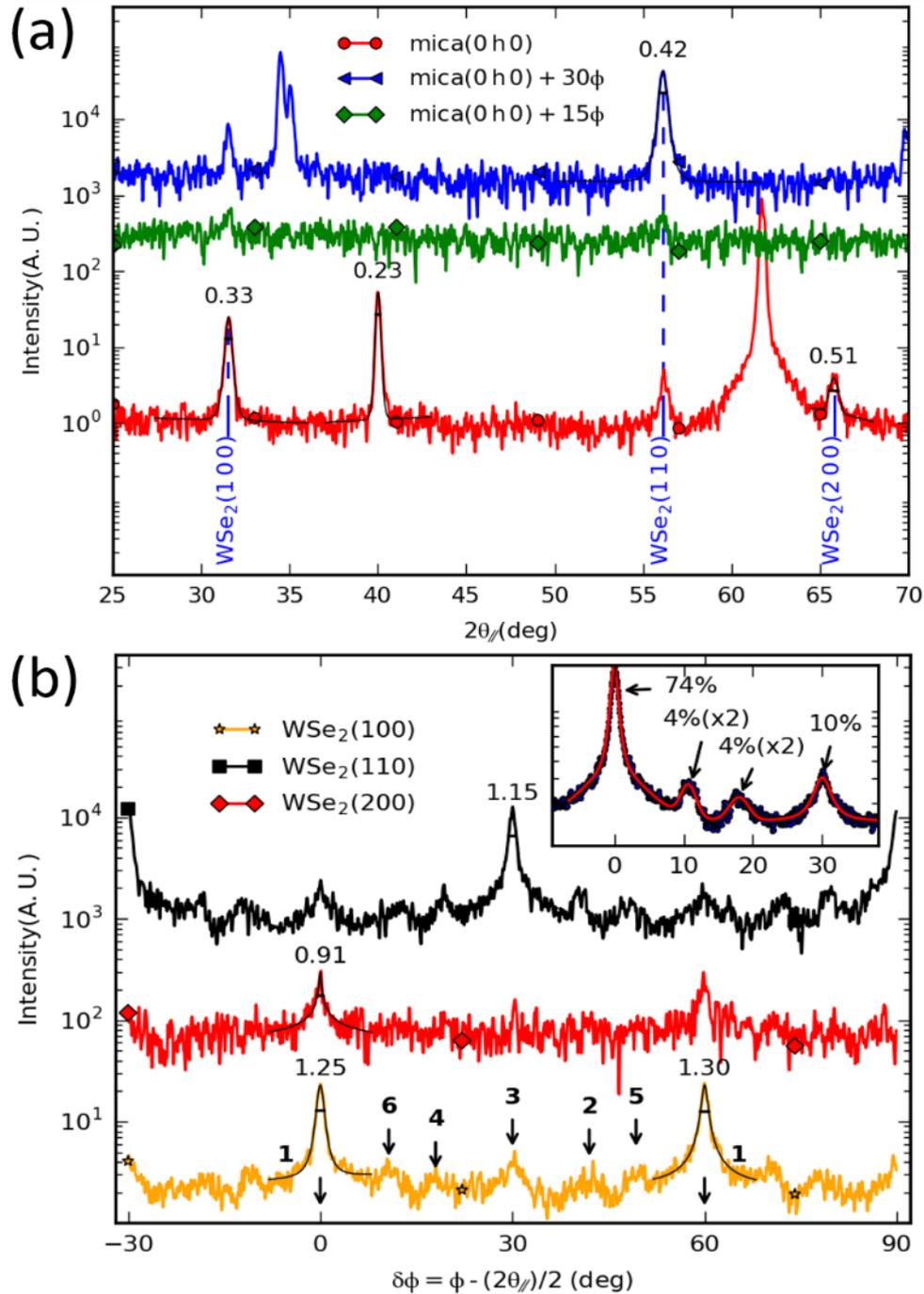

Figure 3: (a) radial and (b) azimuthal grazing incidence x-ray diffraction spectra. The numbers above the Bragg peaks indicate their FWHM. In (a), the non-indexed peaks originate from the substrate. In (b), all the peaks of the WSe$_2$(100) yellow scan are numbered as follows: number **1** corresponds to δϕ=0° and 60°, number **2** to δϕ=42°, number **3** to δϕ=30°, number **4** to δϕ=18°, number **5** to δϕ=49° and number **6** to δϕ=11°. The inset shows the sum of the azimuthal diffraction spectra (100) and (110) using the six-fold and mirror symmetries of WSe$_2$ folded into the 0°-30° angular range (black dots). By fitting the signal intensity (red line), we are able to give the proportion of each peak to the total signal in %.

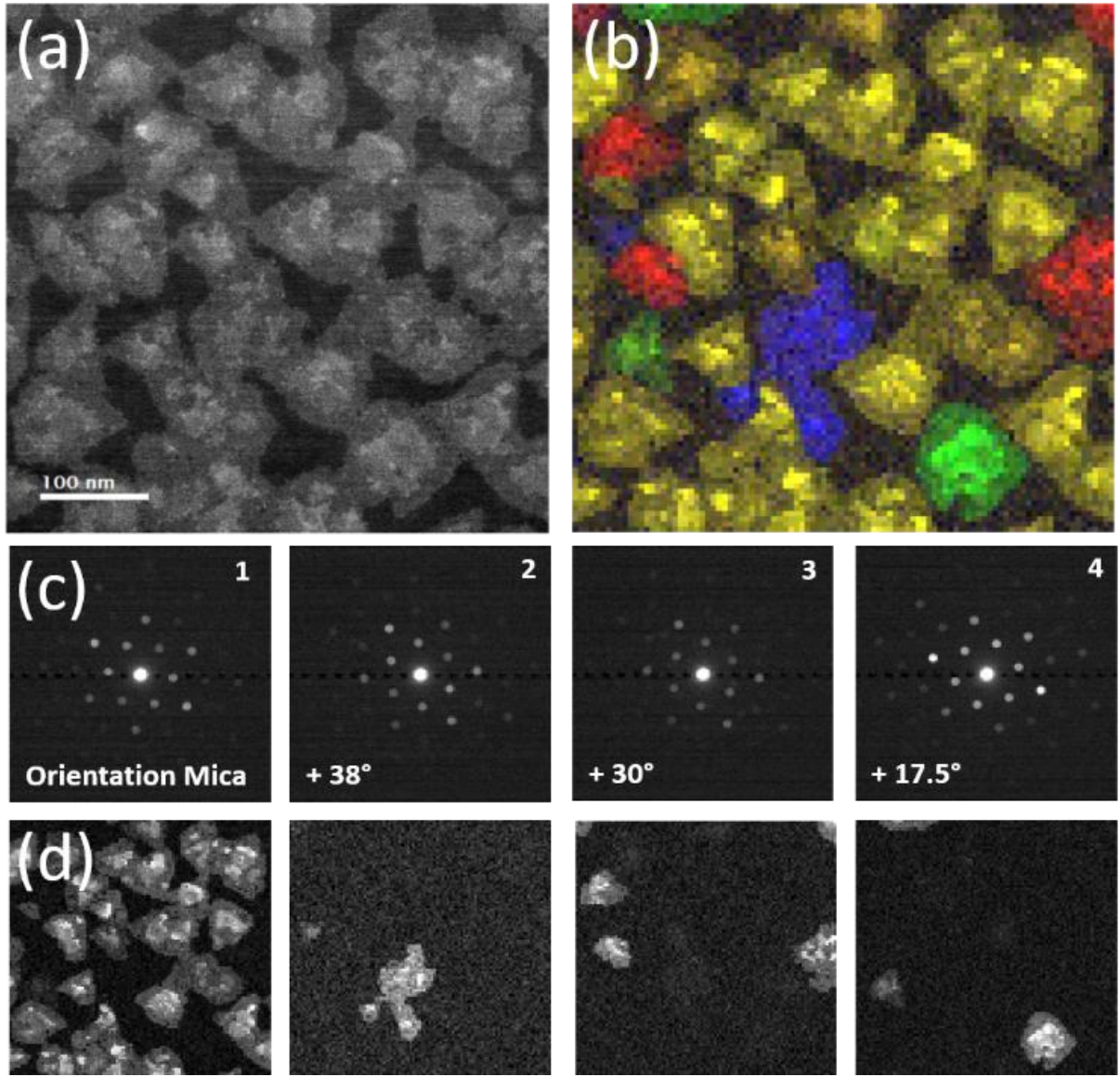

Figure 4: 4D STEM analysis of WSe$_2$ grain orientation. (a) STEM-HAADF image of WSe$_2$ grains. (b) 4D STEM diffraction map showing different grain orientations with a colour code for each, reconstructed by a numerical dark field imaging from different diffraction patterns. The image size is 128×128 pixels. For each pixel, the electron diffraction was recorded in a region of 92×92 pixels in the 500×500 nm image. (c) Diffraction patterns of 4 different crystal directions and (d) corresponding individual dark field images used for the reconstruction of 4D STEM mapping.

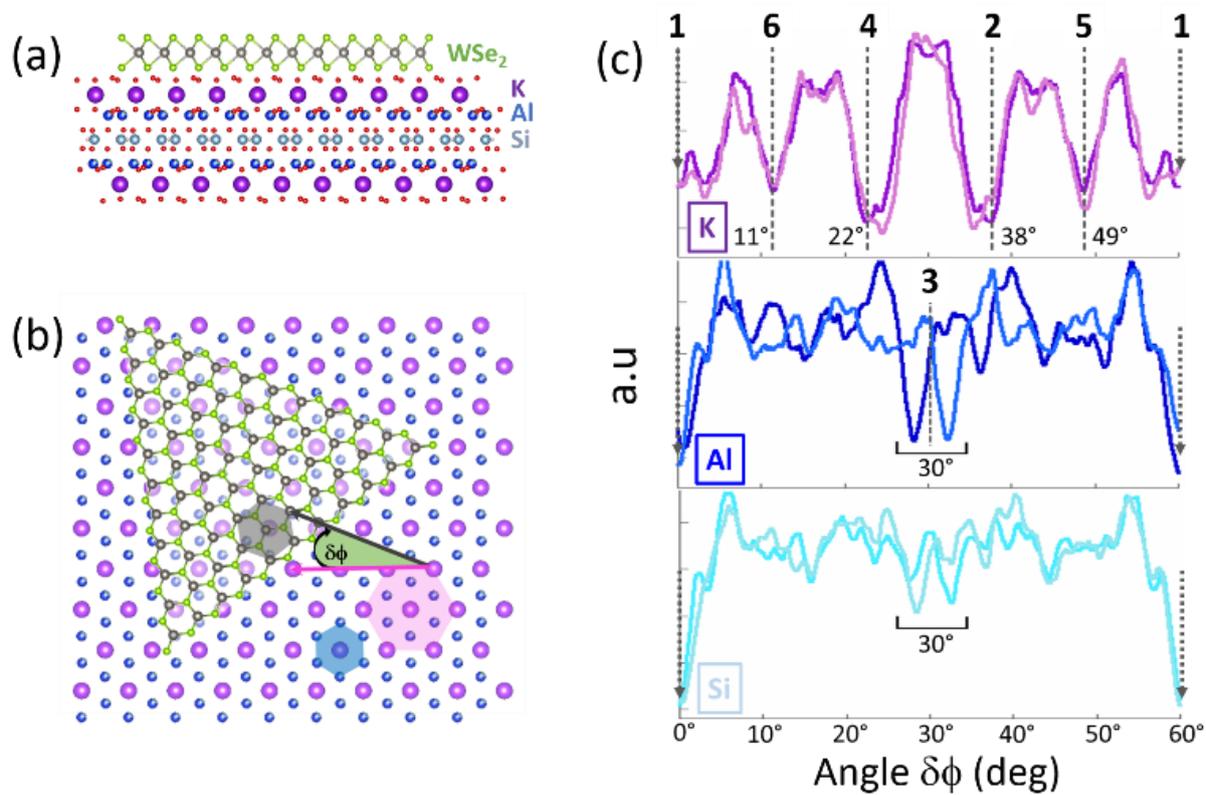

Figure 5: Atomic model in (a) cross-section and (b) top view used for the calculation of atomic coincidences between WSe$_2$ and mica. δϕ is the angle between the WSe$_2$ and mica lattices. (c) Sum of the in-plane distances between each W atom and its first atomic neighbor in the K, Al and Si lattices. For each lattice, the two colors (dark and light) correspond to the 0°-60° and 60°-120° angular ranges respectively. Due to the crystal six-fold symmetry, the two curves should superimpose. However, the slight distortion with respect to the hexagonal lattice due to the actual monoclinic atomic structure of mica leads to a double peak around δϕ=30° for the Al and Si lattices. Those two narrow peaks eventually merge into a single one in x-ray diffraction data due to their finite width. The dips are numbered according to the x-ray diffraction data of Fig. 3b.

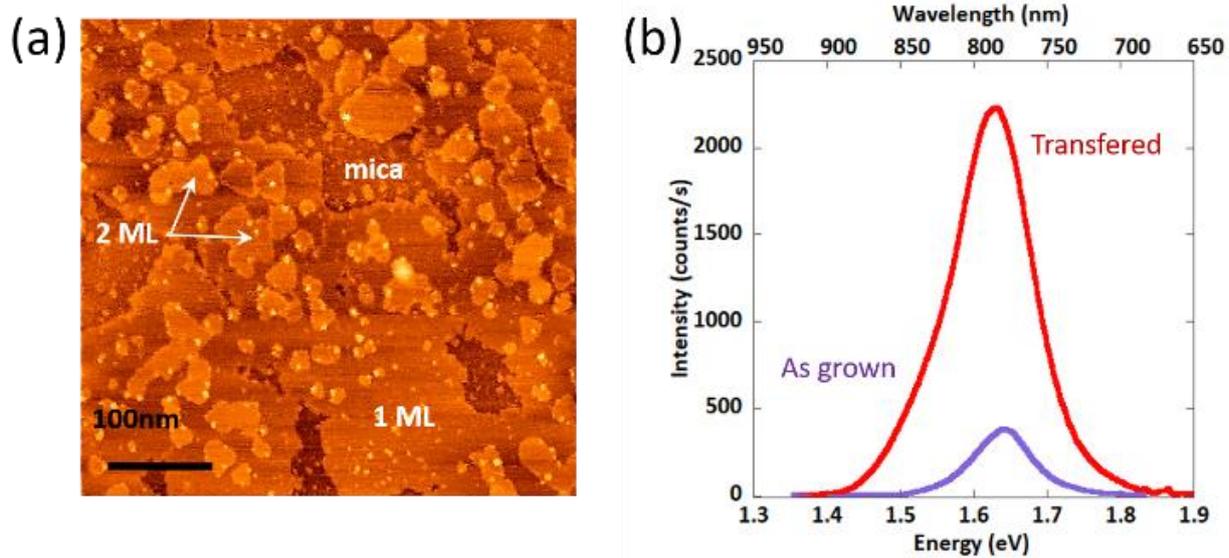

Figure 6: (a) AFM image of one monolayer of WSe$_2$ grown at (T=920°C, F=0.15 Å/min) on mica. (b) Photoluminescence spectra of the as-grown layer on mica and after its transfer on SiN(10nm)/SiO2(300nm)/Si. The laser power and wavelength are 6.2 mW and 532 nm respectively. We use a ×50 objective.